\documentclass[pre,reprint,amsmath,amssymb,showpacs]{revtex4-1}
\usepackage{dcolumn}
\usepackage{bm}
\usepackage[dvips]{color}
\usepackage{subfigure}
\usepackage{graphicx} %eps figures can be used instead
\usepackage{algorithmicx}
\usepackage{pict2e}
\usepackage[ruled]{algorithm}
\usepackage{algpseudocode}
\newcommand{\rvec}{\textit{\textbf{r}} }
\newcommand{\vvec}{\textit{\textbf{v}} }
\newcommand{\evec}{\textit{\textbf{e}} }

\newcommand{\Fvec}{\textit{\textbf{F}} }

\newcommand{\fvec}{\textit{\textbf{f}} }
\newcommand{\Lvec}{\textit{\textbf{L}} }

\newcommand{\Imat}{\textsf{\textbf{I}} }
\newcommand{\Tvec}{\textit{\textbf{T}} }
\newcommand{\Jmat}{\textsf{\textbf{J}} }

\begin{document}

\title{Terminal retrograde turn of rolling rings}

\author{Mir Abbas Jalali$^{1}$}
\email{mjalali@berkeley.edu}
\author{Milad S. Sarebangholi$^2$}
\author{Mohammad-Reza Alam$^{3}$}
\email{reza.alam@berkeley.edu}
\affiliation{
$^1$Department of Astronomy, University of California, Berkeley, California 94720, USA \\
$^2$Department of Mechanical Engineering, Sharif University of Technology, Tehran, Iran \\
$^3$Department of Mechanical Engineering, University of California, Berkeley, California 94720, USA}

%\date{\today}

\begin{abstract}
We report an unexpected reverse spiral turn in the final stage of the motion of rolling rings. 
It is well known that spinning disks rotate in the same direction of their initial spin until they 
stop. While a spinning ring starts its motion with a kinematics similar to disks, i.e. moving along 
a cycloidal path prograde with the direction of its rigid body rotation, the mean trajectory of 
its center of mass later develops an inflection point so that the ring makes a spiral turn and revolves 
in a retrograde direction around a new center. Using high speed imaging and numerical simulations 
of models featuring a rolling rigid body, we show that the hollow geometry of a ring tunes the 
rotational air drag resistance so that the frictional force at the contact point with the ground 
changes its direction at the inflection point and puts the ring on a retrograde spiral trajectory. 
Our findings have potential applications in designing topologically new surface-effect flying 
objects capable of performing complex reorientation and translational maneuvers.
\end{abstract}


\maketitle

%%%%%%%%%%%%%%%%%%%%
%\section{Introduction}

\section{Introduction}
\label{sec:intro}

It is a common experience to spin a coin or a thin disk on a table and observe its rolling 
motion. As the coin keeps rolling, its inclination angle with respect to the table decreases 
while it generates a sound of higher and higher frequency before stopping. According to 
the equations of motion of a rolling rigid body with non-holonomic constraints 
\cite{Borisov02,Borisov03,Borisov13,Kessler02,Ma14,Saux05}, the spin rate must 
diverge to infinity when the disk rests on the table. In real world experiments, however, 
the spin of the disk vanishes within a finite duration of time. Both theoretical and experimental 
studies \cite{Moffatt,Easwar02,Caps04,Leine09,Bildsten02} suggest that the finite life-time of this process 
is due to a combination of air drag and slippage that drain the disk's kinetic energy, but an 
accurate model of dissipative mechanisms is still unknown.

Increasing the thickness of the disk changes the dynamics because of the existence of an 
unstable, inverted-pendulum-like, static equilibrium \cite{Kessler02,Batista06,Shegelski09}. 
Nevertheless, the center of mass of the disk with the global position vector $\rvec_C$ always 
moves on a spiral trajectory \cite{Saux05,Ma14} for low inclination angles, while the orbital 
angular momentum vector $\Lvec=\rvec_G \times \dot \rvec_C$ per unit mass is almost 
aligned with the angular velocity $\boldsymbol{\omega}$ of the disk and we have 
$\Lvec \cdot \boldsymbol{\omega}>0$. Here $\rvec_G$ is the position vector of the 
center of mass with respect to the contact point of the body with the surface. We call this 
spiraling motion a prograde turn. One expects a similar behavior for a ring, but 
experiments reveal a new type of motion, with a retrograde turning phase, which we 
investigate in this paper. 

We present the governing dynamical equations of rolling rings in \S\ref{sec:dynamics}, 
and report experimental and simulation results in \S\ref{sec:experiments} where we 
modify the equations of motion for the effect of air drag, and show how the rolling dynamics
of rings is distinct from disks. The physical origin of retrograde turn is explained 
in \S\ref{sec:phys-origin}. We conclude the paper by remarks on the significance of the 
retrograde turn in rigid-body dynamics, and its analogy with other observed phenomena.

\begin{figure}[b]
\centerline{\hbox{\includegraphics[width=0.24\textwidth]{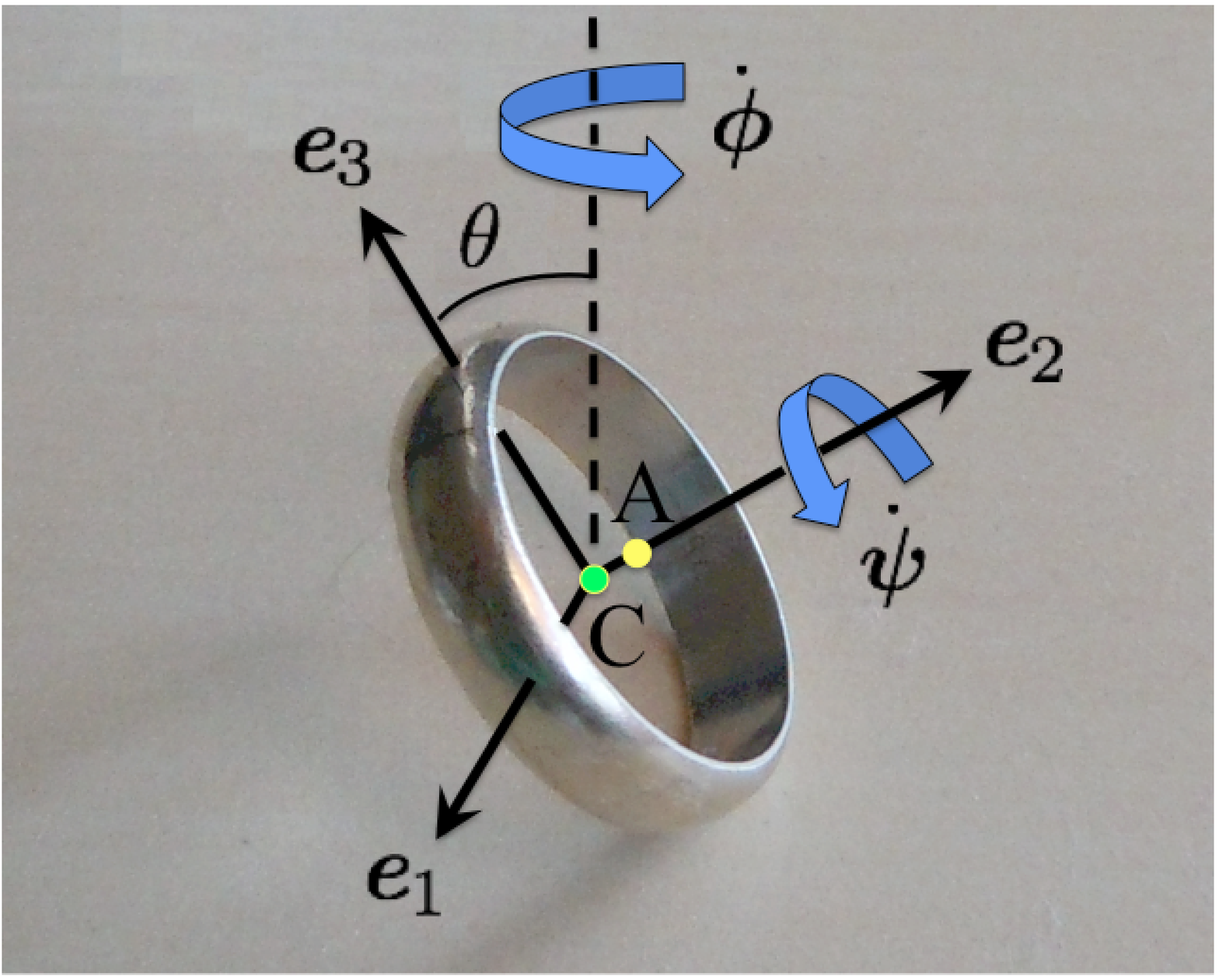} }
\put(-112,90){\textcolor{black}{(a)}}
                    \hbox{\includegraphics[width=0.24\textwidth]{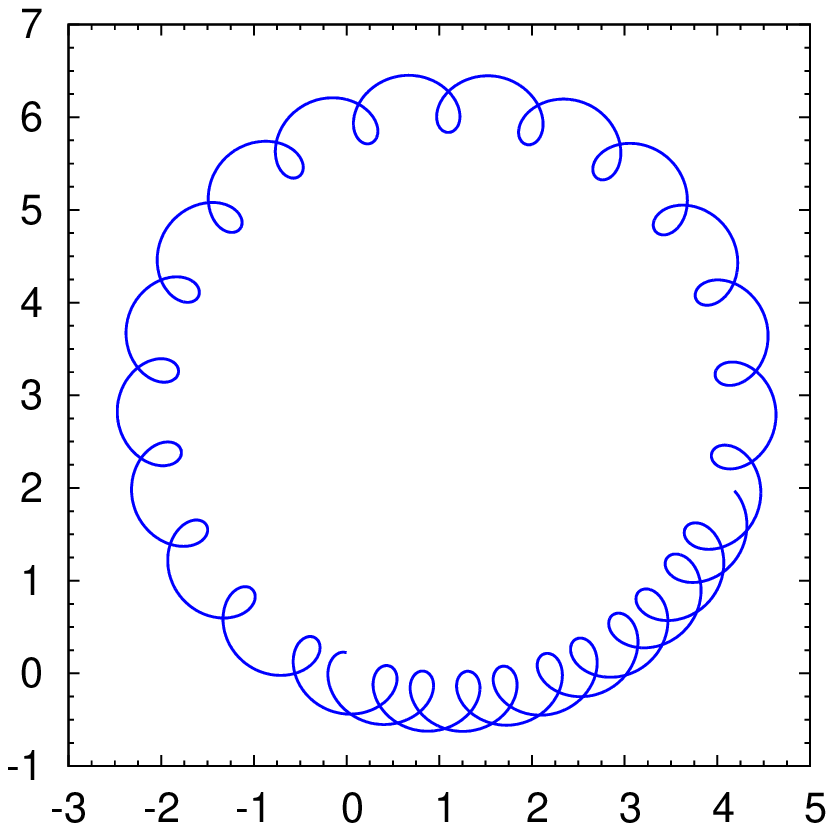} } 
\put(-100,93){\textcolor{black}{(b)}}                    
 }
\caption{(Color online) (a) The geometry of a ring spun on a horizontal table. 
The dashed line is perpendicular to the surface of the table. The origin of the 
coordinate frame defined by $(\evec_1,\evec_2,\evec_3)$ coincides with the 
center of mass of the ring, point $C$. The point $A$ is the center of the upper 
circular edge. (b) The quasi-periodic trajectory of $\rvec_A(t)$ 
projected on the surface of the table, and in the absence of dissipative effects. 
We have set the initial conditions to $\theta_0=0.55$ rad, $\dot\phi_0=4.5$ 
and $\dot\psi_0=0$. All other conditions have been set to zero.}
\label{fig1}
\end{figure}

\section{Dynamics of rolling rings}
\label{sec:dynamics}

We describe the rotation of a ring of the outer radius $R$, width $h$, thickness $w$, and 
mass $m$ by a set of 3-1-2 Euler angles $(\phi,\theta,\psi)$ as shown in Figure \ref{fig1}(a). 
The unit vectors $(\evec_1,\evec_2,\evec_3)$ are along the principal axes of the ring, 
$\evec_1$ is always parallel to the surface of the table, and $\evec_2$ is along the 
symmetry axis of the ring. It is remarked that the ring in Figure \ref{fig1}(a) has not been 
used to quantitatively study the kinematics and dynamics of motion. It is used only for 
the definition of ring geometry, and in supplementary video 1. 
The angular velocity of the ring thus becomes $\boldsymbol{\omega}=\dot \theta \evec_1+(\dot \phi \sin\theta +\dot \psi )\evec_2+\dot \phi \cos \theta \evec_3$. 
We denote the inertia tensor of the ring by $\Imat$ and its angular momentum 
with respect to the center of mass by $\textit{\textbf{L}}_G=\Imat\cdot {\boldsymbol{\omega}}$. 
The equations of the coupled roto-translatory motion thus read 
\begin{eqnarray}
\Imat\cdot \dot{\boldsymbol{\omega}} + \boldsymbol{\Omega} \times \textit{\textbf{L}}_G &=&
-\rvec_G \times \textit{\textbf{F}}, ~~ \rvec_G=(h/2)\evec_2+R \evec_3, \label{eq:rotation} \\
m \ddot \rvec_C &=& \Fvec -mg\left ( \sin\theta \evec_2+\cos\theta \evec_3 \right ), \label{eq:translation}
\end{eqnarray}
where $g$ is the gravitational acceleration, $\rvec_C$ is the global position vector of the center of mass,
$\textit{\textbf{F}}$ is the boundary force at the contact point of the ring and the table, and 
$\boldsymbol{\Omega}=\boldsymbol{\omega}-\dot \psi \evec_2$. Throughout our study we assume that the 
ring is in pure rolling condition and the constraint $\vvec_C=\dot\rvec_C=\boldsymbol{\omega}\times \rvec_G$ 
holds. Equations (\ref{eq:rotation}) and (\ref{eq:translation}) can therefore be combined to obtain the evolutionary 
equations of angular velocities:
\begin{eqnarray}
&{}& \Imat\cdot \dot{\boldsymbol{\omega}} - m\rvec_G \times \left ( \rvec_G\times \dot{\boldsymbol{\omega}} \right )= - \boldsymbol{\Omega} \times \textit{\textbf{L}}_G \nonumber \\
&{}& -m \rvec_G \times \left ( \boldsymbol{\Omega}\times \vvec_C  \right )
 +mg\left [ R \sin\theta - (h/2) \cos\theta \right ]\evec_1.
\label{eq:d-omega}
\end{eqnarray}
It is almost impossible to track the motion of the center of mass experimentally. We therefore 
use the center of the top circular edge of the ring (point $A$ in Fig. \ref{fig1}(a)), with the position 
vector $\rvec_A=\rvec_C+ (h/2)\evec_2$, for measuring the position and velocity of the ring. The 
velocity of point $A$ is related to the speed of the center of mass through 
$\dot\rvec_A=\dot\rvec_C+\boldsymbol{\Omega}\times (h/2)\evec_2$.
We normalize all lengths and position vectors to the mean radius $R-w/2$. Accelerations have been 
normalized so that the initial value of $R\dot\phi^2$ at $t=0$ equals the experimental value 
$\approx 20.2 g$. 

Integration of equation (\ref{eq:d-omega}) for initial conditions $\dot\theta_0=0$ and 
$\theta_0 > \arctan[h/(2R)]$ show that the center of mass of the ring moves on a generally 
quasi-periodic cycloidal orbit. A typical quasi-periodic orbit is shown in 
Fig. \ref{fig1}(b) for $R=1.025$, $h=0.88$ and $w=0.05$, which correspond to the ring in 
our experiments discussed below. The size of the inner turning loop of cycloids is a 
function of $\dot\psi_0/\dot\phi_0$ and $\theta_0$. Such orbits, however, are not observed 
in real world experiments. Spinning a wedding ring on a glass or wooden table shows 
that the motion is composed of two prominent phases. In the first phase, the ring spins 
and travels similar to the prograde turn of a coin/disk, but in contrast with a disk that 
continues prograde spiraling until its resting position, it abruptly makes a retrograde 
spiral turn before stopping (supplementary video 1). 
The retrograde turn does not belong to the phase space structure of equation (\ref{eq:d-omega}), 
nor is it observed in spinning disks. 

\section{Experimental results and theoretical simulations}
\label{sec:experiments}

\begin{figure}
\begin{center}
\includegraphics[width=0.23\textwidth]{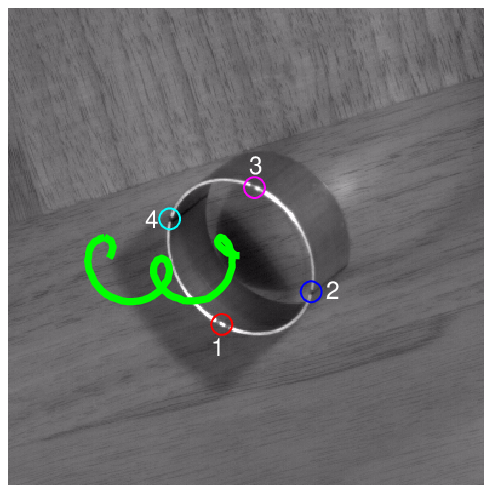}
\put(-100,95){\textcolor{white}{(\bf a)}}
\hspace{0.01in}
\includegraphics[width=0.23\textwidth]{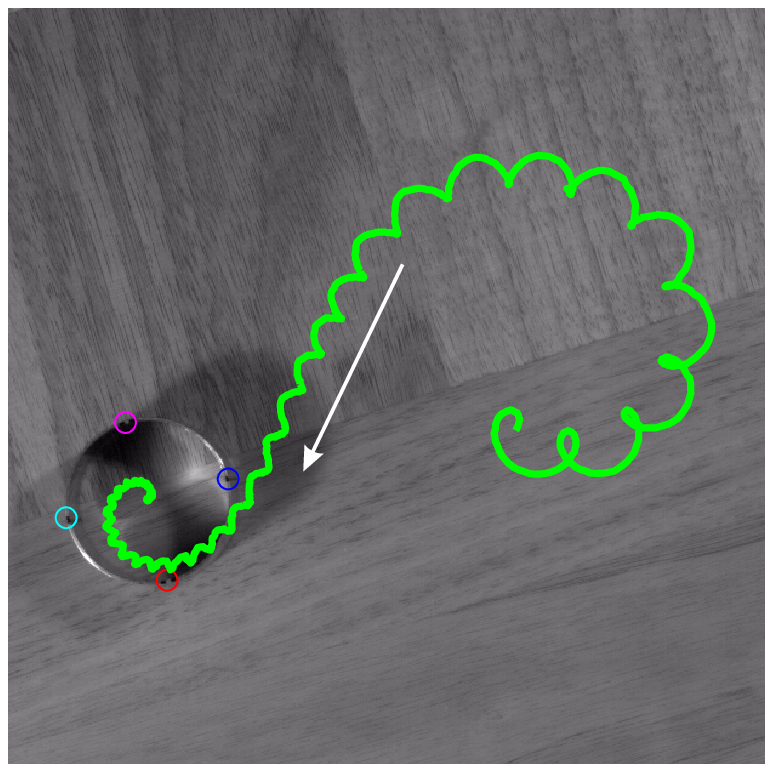}
\put(-100,95){\textcolor{white}{(\bf b)}}
\end{center}
\begin{center}
\includegraphics[width=0.228\textwidth]{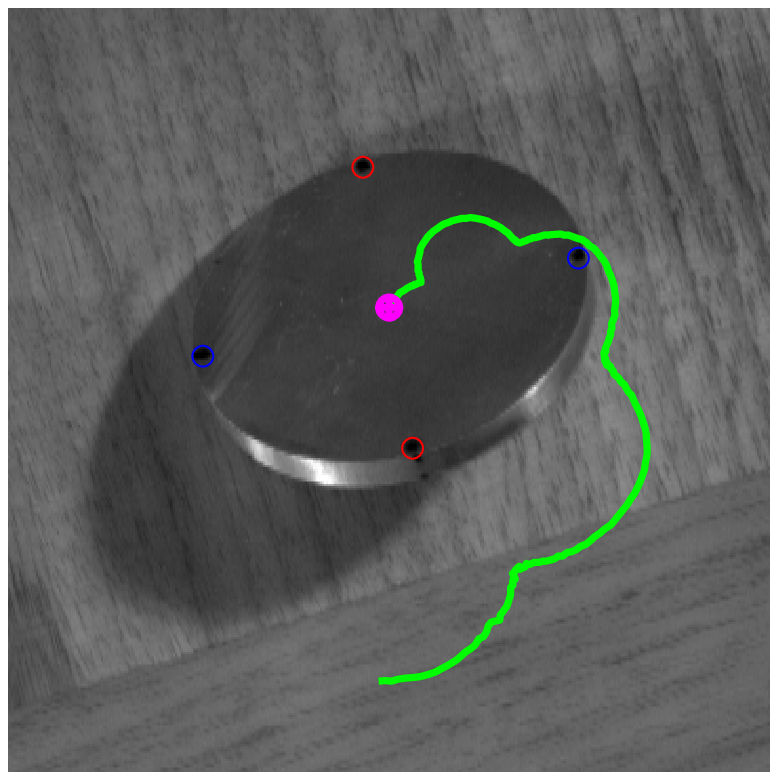}
\put(-100,95){\textcolor{white}{(\bf c)}}
\hspace{0.01in}
\includegraphics[width=0.23\textwidth]{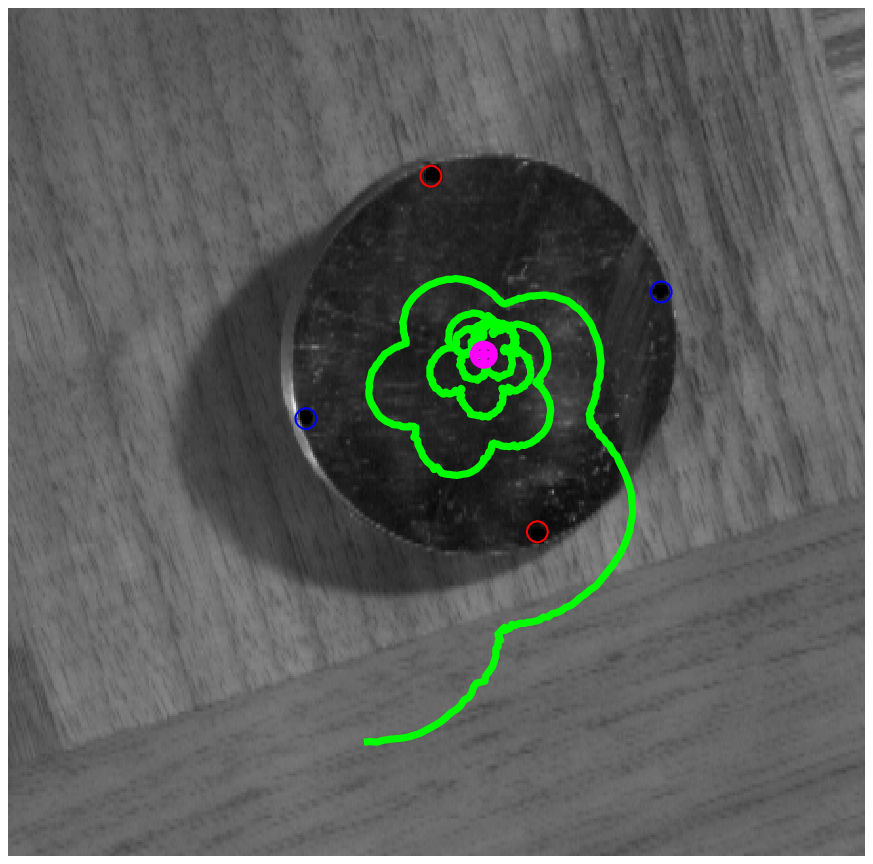}
\put(-100,95){\textcolor{white}{(\bf d)}}
\end{center}
\vspace{-0.2in}
\caption{(Color online) (a) The first cycloidal turn of the 
ring. The four marks for motion tracking 
have been labeled by numbers 1, 2, 3 and 4. (b) The full trajectory of 
the center of the top circular edge, point $A$, of the ring. The ring undergoes 
a directional walk along the arrow. (c) The first few prograde turns of the 
rolling disk. (d) The trajectory of the center of mass of the disk until it reaches
a small inclination angle.}
\label{fig2}
\end{figure}

To understand the ring dynamics, we prepared a high-speed imaging set-up and 
spun a ring of $R=20.66 \, {\rm mm}$, $w=1\, {\rm mm}$ and $h=18 \, {\rm mm}$ on a 
polished and waxed wooden table. The ring has been cut from a steel tube with 
circular cross section. We rotated and released the ring by hand, but assured that the initial 
conditions satisfy $\dot\theta_0 \approx 0$ and $\theta_0 > \arctan[h/(2R)]$. 
To trace the translational and rotational motions, we put four marks in a cross configuration 
at the top circular edge of the ring, and stored their coordinates (in pixels) while filming 
the motion (Fig. \ref{fig2}(a)). The centroid of these marks has the position vector $\rvec_A$. 
Figs 2a,b and supplementary video 2 show the projection of the trajectory of $\rvec_A(t)$ 
on the surface of the table for one of our experiments. The Euler angles $\theta$ and $\psi$ 
can be computed from the formulae
$1+\sin^2(\theta)$=$\left (L_{13}^2+L_{24}^2 \right )/D^2$ and 
$1+\sin^2(\psi)\left [\sin^2(\theta) -1 \right ]$=$L_{13}^2/D^2$
where $L_{13}$ and $L_{24}$ are the apparent distances between the points 1 and 3, 
and 2 and 4, respectively, and $D=2R-w$ is the mean diameter of the ring. Our experimental 
error level in computing $\rvec_A(t)$ has been $\approx 5\%$ because of image distortions. 
There are two reasons behind image distortions: perspective effects and barrel distortions
(the field of view of the lens is bigger than the CCD size). Perspective distortions are functions 
of (i) the distance of the ring from the line of sight of the camera, and (ii) the Euler angles. 
The mean error threshold due to all these effects is roughly the measured value of 
$1-(L_{13}+L_{24})/(2D)$ after the stopping of the ring.

The trajectories of point $A$ displayed in Figs \ref{fig2}(a,b) and supplementary 
video 2 unveil unique features of the ring's motion. An initial prograde turning phase 
occurs along cycloidal curves similar to what we observe in Fig. \ref{fig1}(b). As time 
elapses, the inner turning loops of cycloids shrink and evolve to cuspy turning points 
that connect half-circle-shape arcs. The radii of half-circle steps decrease and the 
motion becomes directional along the arrow until a retrograde spiral turn begins at 
an inflection point. 

To better distinguish the differences between the trajectories of rings and disks, we 
repeated our experiment for an aluminum disk of diameter $D=63.5$ mm and width  
$h=6.14$ mm, and recorded its trajectory. Figs \ref{fig2}(c,d) and the supplementary video 3 
show the inspiraling motion of the disk's center of mass. This is a generic behavior of
rolling disks, regardless of their thickness \cite{Ma14}.

\begin{figure}
\begin{center}
\includegraphics[width=0.46\textwidth]{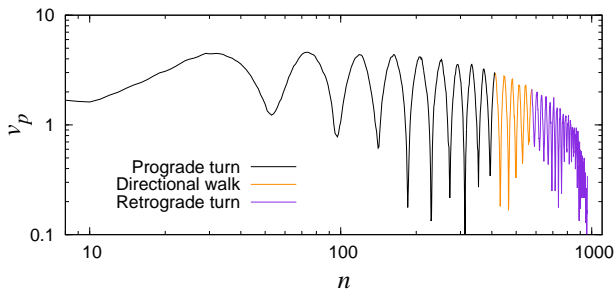}
\end{center}
\vspace{-0.25in}
\caption{(Color online) The experimentally computed velocity of the ring parallel to 
the surface of the table. The magnitude of $\vvec_p$ in units of pixel/frame 
versus the frame number $n$. The imaging speed has been 300 frames per second, 
and approximately $135.6$ pixels correspond to $40.3$ mm.}
\label{fig3}
\end{figure}

Using the coordinates of the four markers on the ring, we have computed the magnitude of the velocity 
$\vvec_p=\vvec_A-(\vvec_{A} \cdot \evec_{\perp})\evec_{\perp}$, which is parallel to the surface 
of the table, and plotted it in Fig. \ref{fig3} versus the frame number $n$. 
Here $\evec_{\perp}=\sin(\theta) \evec_2+\cos (\theta) \evec_3$ 
is the unit vector normal to the surface. At the highest ($\theta=\theta_{\rm min}$) and lowest 
($\theta=\theta_{\rm max}$) vertical positions of the center of mass, $\vvec_p$ becomes identical 
to $\vvec_A$. The envelope of the velocity profile has a shallow decline up to and after the 
retrograde turn, followed by a steep fall and termination of the motion. 

We have repeated our experiments with rings of different $h/R$ ratios and observed the 
retrograde turn in all cases. The spiral turn is more prominent for $h/R \approx 1$ as in 
the ring of Fig. \ref{fig2}. Several mechanisms like 
rolling friction, slippage \cite{Kessler02}, air drag \cite{Moffatt}, and even elastic vibrations \cite{Villanueva05} 
can be held responsible for the phenomenon. Our numerical calculations in \S\ref{sec:phys-origin} show that 
the normal contact force multiplied by the coefficient of friction never exceeds the lateral frictional force, 
and therefore, slippage does not play any role in the occurrence of the retrograde spiral turn. Moreover, 
elastic vibrations may change the course of motion only if their frequencies resonate with the precession 
frequency $\dot\phi$ of the ring. We have not observed any signs of resonances in the signals of $v_p$ 
and $\dot\phi$. It is shown that including only the air drag fully captures the physics of the retrograde 
turn. In the presence of external drag torques, equation (\ref{eq:d-omega}) takes the form
\begin{eqnarray}
\Jmat \cdot \dot{\boldsymbol{\omega}} &=& \fvec \left (\boldsymbol{\omega},\theta \right )+\Tvec_{\rm drag},
\label{eq:short-d-omega} \\
\fvec &=& - \boldsymbol{\Omega} \times \textit{\textbf{L}}_G -
m \rvec_G \times \left ( \boldsymbol{\Omega}\times \vvec_C  \right ) \nonumber \\
&{}& +mg\left [ R \sin\theta - (h/2) \cos\theta \right ]\evec_1, \nonumber \\
\Jmat &=& \Imat + m 
\left [ 
\begin{array}{ccc}
h^2/4+R^2  & 0  & 0  \\
0  & R^2  & -hR/2  \\
0  & -hR/2  & h^2/4  
\end{array}
\right ]. \nonumber 
\end{eqnarray}
where $\Jmat$ is a constant matrix, $\fvec$ is a vector function of the angular velocity 
$\boldsymbol{\omega}=\boldsymbol{\Omega}+\dot \psi \evec_2$ and the Euler angle $\theta$, and $\Tvec_{\rm drag}$ is the resultant 
drag-induced torque. 
The exact value of drag force on a general bluff body undergoing a three dimensional motion 
is very difficult to calculate, and is not available. In fact, the behavior of the viscous drag is so 
complicated that even for basic symmetric two dimensional objects under uniform transnational 
motion we need to entirely rely on empirical formulae \cite{SG00}. If the bluff body in rotation is 
symmetric, then in order to estimate the drag moment the best approximation is to use the 
rotational drag coefficient and implement it on the net angular velocity vector \cite{I03}. 
This gives a drag moment vector in the same direction as of the angular velocity vector. 

If the bluff object is not symmetric, then it is clearly not possible to define a single rotational 
drag coefficient for the general three-axes rotations. For a general three dimensional object, 
every direction of the angular velocity corresponds to a different rotational drag coefficient 
that needs to be found empirically. Here and as an approximation, we assume that the 
vector $\Jmat^{-1}\cdot \Tvec_{\rm drag}$ is proportional to $\boldsymbol{\omega}$.
This means rotation about a given axis does not induce angular acceleration about 
other axes. The rational comes from the observation that releasing the ring from a stationary 
initial condition with $\theta >0$ and $\boldsymbol{\omega}=\textbf{0}$ yields a simple 
accelerating rotation about the unit vector $\evec_1$ until the ring hits the ground. Therefore, 
the air drag does not couple $\omega_2$ and $\omega_3$ 
to $\omega_1$. Moreover, the drag force corresponding to a pure rotation 
about $\evec_3$ does not affect $\omega_2$ and $\omega_1$ when 
$\theta \rightarrow \pi/2$. The main approximation made here is for rotation 
about $\evec_2$: as the ring rotates about $\evec_2$ and undergoes a 
translational motion along $\evec_1$ due to rolling constraint, 
even a small-amplitude rotation about $\evec_3$ couples drag 
force components. Finding a more accurate model for $\Jmat^{-1}\cdot \Tvec_{\rm drag}$ 
is beyond the scope of this study. We are not aware of any systematic method to 
experimentally determine drag force components near a boundary. The only reliable 
way is to use computational fluid dynamics (CFD) methods, which can be considered 
as potentially interesting problems for future works. Below it is shown that even 
our approximate model captures the physics of the problem very well.

We define the three rotational drag coefficients $C_i$ ($1,2,3$) corresponding 
to the three major axes of the ring and write:
\begin{eqnarray}
\boldsymbol{\tau} \equiv \Jmat^{-1}\cdot \Tvec_{\rm drag} = - \sum_{i=1}^{3} C_{i} \, \vert \omega_i \vert \, \omega_i \evec_i, 
~~\omega_i=\boldsymbol{\omega} \cdot \evec_i.
\label{eq:rotation-drag}
\end{eqnarray}
Variants of this approach are used in naval hydrodynamics \cite{R12}, flight dynamics \cite{C02} 
and low Reynolds number swimming \cite{Ch08}. The rotational drag coefficients $C_{i}$ implicitly 
depend on the Reynolds number $Re$ and the reference area of the ring exposed to airflow. If the ring was far from 
any wall/surface, the rotational symmetry about the $\evec_2$-axis would imply $C_{1}=C_{3}$,
but for rolling rings this identity does not necessarily hold. Let us define the Reynolds number as $Re=2\vert \dot \rvec_C\vert R/\nu_a$ where $\nu_a$ in the kinematic viscosity of the air. According to the velocity data of Fig. \ref{fig3}, the Reynolds number satisfies $Re \lesssim 800$. 

\begin{figure}
\centerline{\hbox{\includegraphics[width=0.24\textwidth]{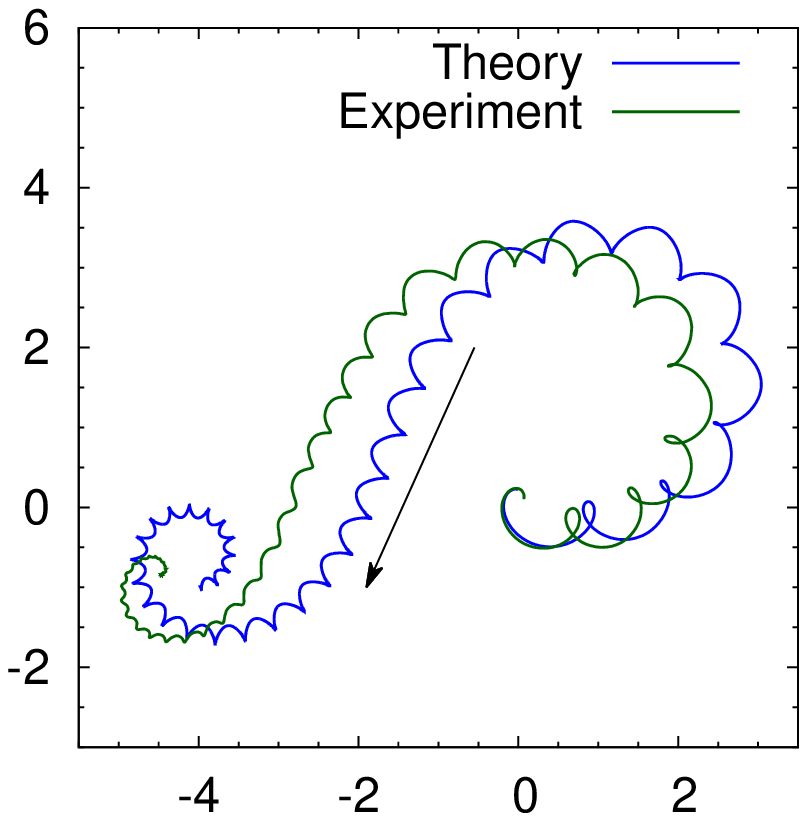} }
\put(-100,97){\textcolor{black}{(\bf a)}}
                    \hbox{\includegraphics[width=0.24\textwidth]{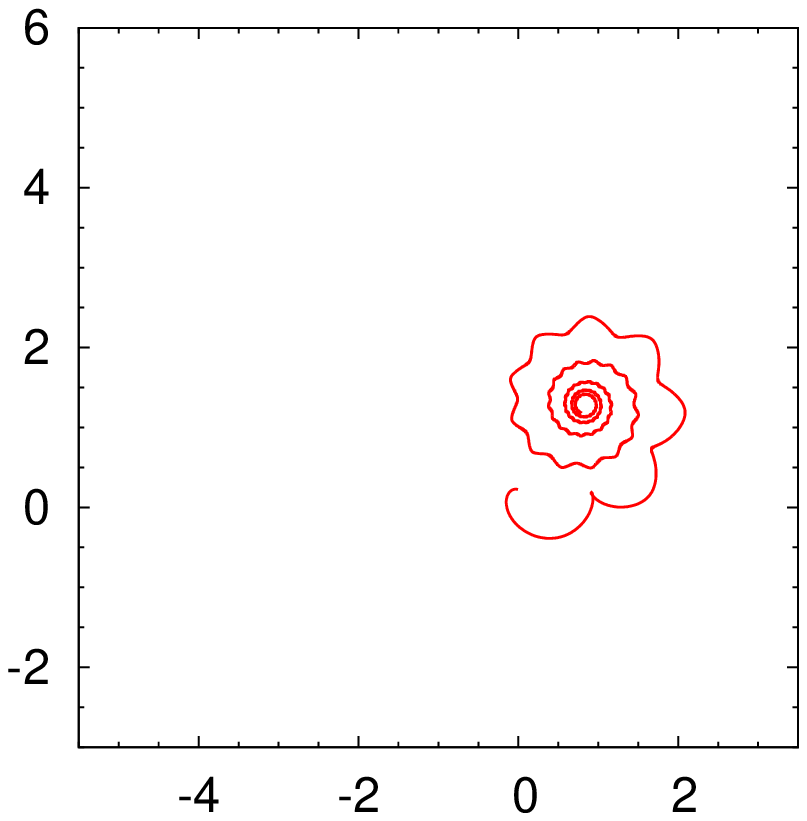} } 
\put(-100,97){\textcolor{black}{(\bf b)}}                    
 }
 \vspace{0.1in}
\centerline{\hbox{\includegraphics[width=0.46\textwidth]{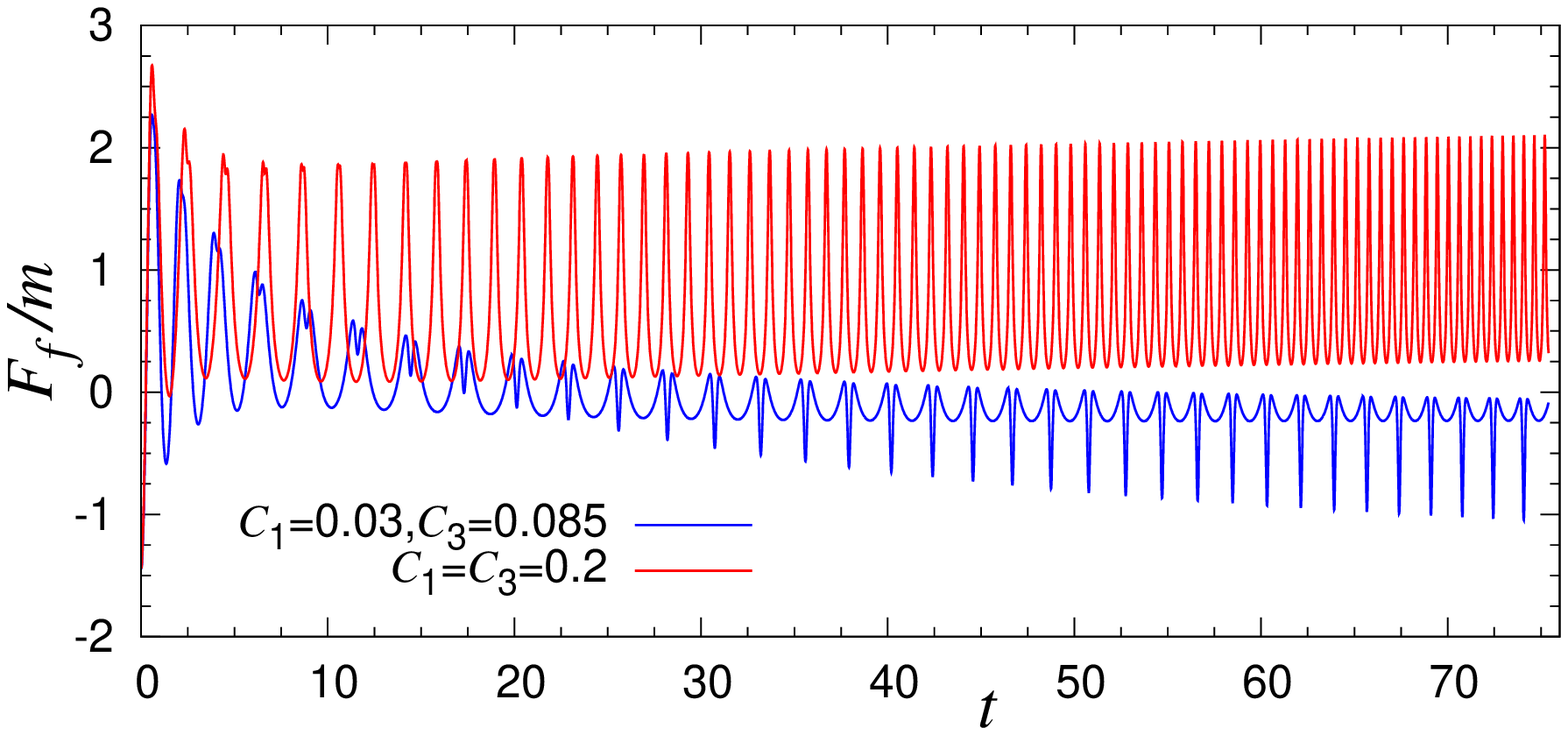} }
\put(-195,85){\textcolor{black}{(\bf c)}}
 }
\caption{(Color online) (a) The projection of the 
simulated trajectory of $\rvec_A(t)$ on the surface for the same ring of 
Figs \ref{fig1}(a,b) and with $(C_1,C_2,C_3)=(0.03,0.063,0.085)$. 
We have also reproduced the experimental trajectory of Fig. \ref{fig2}(b) for 
comparison. The motion between the initial prograde and final retrograde 
turning phases is directional along the arrow. (b) Same as panel (a) 
but for $(C_1,C_2,C_3)=(0.2,0.063,0.2)$. All lengths and position vectors 
have been normalized to the mean radius $R-w/2$ of the ring. (c) The 
variation of the friction force $F_f$ versus time for the models of panels 
(a) and (b). Variables are in dimensionless units.}
\label{fig4}
\end{figure}

Equations (\ref{eq:short-d-omega}) and (\ref{eq:rotation-drag}) yield
$\dot{\boldsymbol{\omega}}=\Jmat ^{-1} \cdot \fvec \left (\boldsymbol{\omega},\theta \right )+\boldsymbol{\tau}$.
We numerically integrate this equation using the initial conditions that we measure 
at the first inner turning point of Fig. \ref{fig2}(b). In that specific position, 
the angular velocity $\dot\theta$ vanishes, and we find $\theta_{\rm min} \approx 0.55$ rad, 
$\phi \approx -0.38$ rad, $\dot\psi \approx 0$, and $\dot\phi = v_A/[R\sin(\theta_{\rm min}) - (h/2)\cos(\theta_{\rm min})] \approx 4.5$. The computed initial angular velocities are dimensionless. Without loss of generality, 
we assume $\psi(0)=0$. The initial velocity of the center of mass is calculated using the rolling condition. 
To the best of our knowledge, the drag coefficients of a ring have not been measured or tabulated so 
far. Therefore, we constrain the parameter space $(C_{1},C_{2},C_{3})$ by generating all orbits that 
resemble the experimental trajectory displayed in Fig. \ref{fig2}(b). We find the best match between 
theoretical and experimental trajectories by setting $C_{1} \approx 0.03$, $C_2 \approx 0.063$, 
and $C_{3} \approx 0.085$. The projection of 
the simulated trajectory of $\rvec_A(t)$ on the surface has been demonstrated in Fig. \ref{fig4}(a)
together with the experimental trajectory. According to our computations, 
the topology of the trajectory is not sensitive to the variations of $C_1$ over the range 
$0.01 \lesssim C_1 \lesssim 0.1$ when the quotients $C_1/C_2$ and $C_1/C_3$ are kept constant. 
By varying $C_1$ We observe only minor differences in the location and size of the terminal spiral 
feature.

The actual and simulated trajectories are similar in many aspects, including 9 and 15 cycles 
that they make, respectively, before the directional walk and retrograde turn phases. Their major 
differences are the long-lived last spiral stage of the simulated trajectory, and a drift. We suspect 
that the observed drift has been due to (i) uncertainties in calculating the initial angular velocities 
through the de-projection of the images and (ii) slippage at some cuspy turning points that has 
slightly changed the direction of $\vvec_A$. For the existing discrepancy in the final spiral 
path we have the following explanation: as the motion of the ring slows down, $Re$ decreases 
and the drag coefficients increase. Consequently, the life-time of the spiral 
turn is shorter in reality. We would expect a better match with the experiment if the accurate 
profiles of the drag coefficients were known in terms of $Re$. We have repeated our experiments 
on glass sheets and polished steel plates, and obtained similar results. Therefore, deformation 
of the surface does not play a decisive role in the onset of retrograde turn. 

\section{Physical origin of retrograde turn}
\label{sec:phys-origin}

A fundamental question is why disks do not make 
a retrograde turn like rings? This returns to differences in their aerodynamic properties 
near the ground: air can always 
flow through the central hole of the ring, with the drag force components $C_i$ ($i=1,2,3$) 
in all directions coming mostly from the skin friction scaled by ${\cal O}(Re^{-1/2})$ in the laminar 
flow conditions (with $Re\lesssim 800$) of our experiments \cite{Sh89}. For disks, however, air is 
trapped and compressed between the disk and the ground, the contribution of the \textit{form drag} 
to $C_1$ and $C_3$ is significant, and the drag coefficients $C_1$ and $C_3 \sim {\cal O}(1)$ are (almost) 
independent of the Reynolds number when $Re>100$ \cite{Sh89}. Therefore, for rolling disks we 
expect $C_1/C_2 \gg 1$ and $C_3/C_2 \gg 1$. By taking the same initial conditions for the ring in our 
experiments, we used $C_2=0.063$ and $C_1=C_3=0.2$, and found that the corresponding 
simulated trajectory of $\rvec_A(t)$ (Fig. \ref{fig4}(b)) is a single prograde spiral analogous 
to the experimentally measured trajectory of Fig. \ref{fig2}(d). This shows the role of enhanced 
drag torque about the diameter in maintaining the prograde turn. 

We have found that the evolution of the lateral component 
$F_f=\Fvec\cdot [\cos(\theta)\evec_{2}-\sin(\theta)\evec_3]$ of the frictional force at 
the contact point is the dynamical origin of the retrograde turn. The ring maintains its 
motion on a trajectory as in Figs \ref{fig1}(b) and \ref{fig2}(d) if the 
lateral force satisfies $F_f > 0$ and supports the centrifugal acceleration needed for 
the prograde turn, especially when the center of mass passes through its lowest 
vertical position (with $\theta=\theta_{\rm max}$ and $\dot\theta=0$) at each 
cycle. At this point, the kinetic energy of the center of mass is maximum and its potential 
energy takes a minimum. We remark that the component $F_1=\Fvec \cdot \evec_1$ of 
$\Fvec$ is also caused by friction, but it helps the rolling and 
cannot balance the centrifugal acceleration at turning points. Our computations (Fig. \ref{fig4}(c)) 
show that because of drag torques, a local minimum that develops on the profile of $F_f$ 
at $\theta_{\rm max}$ gradually becomes spiky and flips sign from positive to negative. 
As the ring experiences the strong negative kicks of $F_f$, the centrifugal acceleration switches 
sign as well, and the ring starts to revolve around a new point by retrograde turning. 
This process does not happen for disks, for $\omega_1$ and $\omega_3$ decay 
quickly due to a large $C_1$ and $C_3$, and the orbital angular momentum 
$\rvec_G \times \dot\rvec_C$ is dominated by the $\evec_2$-component. 
Consequently, $F_f$ that supports the centrifugal acceleration remains positive as 
$\theta \rightarrow \pi/2$ (Fig. \ref{fig4}(c)). The coefficient of static friction for the 
surface on which we had spun our ring was $\mu\approx 0.4$. We computed the 
normal component of the contact force $F_N=\Fvec\cdot \evec_{\perp}$ over the 
entire motion of the ring and found that the inequality $\mu F_N>[F_1^2+F_f^2]^{1/2}$ 
holds at all turning points with $\theta=\theta_{\rm max}$. 
Therefore, slippage is not expected to play any major role in the qualitative features 
of the motion.

In summary, the aerodynamic interactions of spinning bodies can lead to complex, 
and sometimes unpredictable, results depending on the shape of the object and the 
initial conditions of its motion. Three well-known examples of spinning objects that 
significantly change their course of motion are the returning boomerang, soccer balls,
and frisbees that fly along curved paths. Neither a boomerang nor a frisbee can 
move on curved trajectories without aerodynamic effects. Our finding for spinning 
rings is a new case where the frictional force and aerodynamic forces near the 
surface collaborate to change the course of motion.

\end{document}